\documentclass[twocolumn,showpacs,preprintnumbers,amsmath,amssymb,prb]{revtex4} \newcommand\PrePrint{0}
\usepackage[dvips]{color,graphics,epsfig,rotating}
\usepackage{graphicx}
\usepackage{dcolumn}
\usepackage{bm}
\usepackage{color}
\usepackage{ifthen}

\makeatletter
\ifthenelse{\equal{\PrePrint}{1}}
{
  \def\@dotsep{4.5}
}
{
}
\makeatother

\newcommand\eql[2] 
{
\begin{equation}\label{#1}
\begin{split}
#2
\end{split}
\end{equation}
}

\newcommand\ket[1]     {|{{#1}}\rangle}

\newcommand\Hop        {{\hat{H}}}

\newcommand\Sop        {{\hat{S}}}

\newcommand\vop        {{\hat{v}}}

\newcommand\GAUSSIAN[1][]{{\footnotesize{GAUSSIAN#1}}}
\newcommand\GAMESS     {{\footnotesize{GAMESS}}}
\newcommand\NWCHEM     {{\footnotesize{NWCHEM}}}

\newcommand\De         {D_e}
\newcommand\mEh        {\textrm{m}\ensuremath{E_\textrm{h}}}

\newcommand\HzO        {H$_{\textrm{2}}$O}
\newcommand\Fz         {F$_{\textrm{2}}$}

\definecolor{Green}{rgb}{0.2,0.96,0.2}
\definecolor{Remarks}{rgb}{1,0.3,0.3}
\definecolor{Extra}{rgb}{0.2,0.2,1}
\definecolor{Blue}{rgb}{0.2,0.3,1}
\definecolor{Black}{rgb}{0,0,0}

\newcommand\FN[1][] {\footnotemark[#1]}

\newcommand\COMMENTED[1] {}
\newcommand\REMARKS[1]   {\textcolor{Remarks}{\textbf{#1}}}
\renewcommand\REMARKS[1] {} 
\newcommand\FIGDIR[1]   {figs/}

\begin{document}

\title{
Eliminating spin contamination in
auxiliary-field quantum Monte Carlo:\\
realistic potential energy curve of F$_2$
}

\author{Wirawan Purwanto}
\author{W. A. Al-Saidi}
\altaffiliation[Present address: ]{Department of Physics, Cornell University,
Ithaca, New York 14853, USA}
\author{Henry Krakauer}
\author{Shiwei Zhang}
\affiliation{Department of Physics, College of William and Mary,
Williamsburg, Virginia 23187-8795, USA}

\date{\today}

\begin{abstract}

The use of an approximate reference state wave function $\ket{\Phi_r}$ in
electronic many-body methods can break the spin symmetry of
Born-Oppenheimer spin-independent Hamiltonians. This can result in
significant errors, 
especially when bonds are
stretched or broken. A simple spin-projection method is introduced for auxiliary-field
quantum Monte Carlo (AFQMC) calculations, which yields
spin-contamination-free results, even with a spin-contaminated
$\ket{\Phi_r}$. The method is applied to the difficult \Fz{} molecule,
which is unbound within unrestricted Hartree-Fock (UHF).
With a UHF $\ket{\Phi_r}$,
spin contamination causes large systematic
errors and long equilibration times in AFQMC in the intermediate, bond-breaking region.
The spin-projection method eliminates these problems, and delivers
an accurate potential energy curve from equilibrium to the dissociation limit
using the UHF $\ket{\Phi_r}$.
Realistic potential energy curves
are obtained with a cc-pVQZ basis.
The calculated spectroscopic
constants are in excellent agreement with experiment.
\end{abstract}

\COMMENTED{
\begin{verbatim}
---------------------------------------------------------------------------
GAFQMC PAPER 1: F2 MOLECULE/SPIN CONTAMINATION (preprint edition)
CVS $Id: F2-paper.tex,v 1.19 2007-12-11 16:04:47 wirawan Exp $
---------------------------------------------------------------------------



\end{verbatim}
}



\maketitle

\section{Introduction}

A standard approach in many-body electronic structure methods is to
obtain ground and excited state energies from an approximate reference
state wave function $\ket{\Phi_r}$.
For example, the coupled-cluster (CC) approximation with
single, double, and perturbative triple excitations [CCSD(T)],
which is widely available in
quantum chemistry computer codes, typically uses
a Hartree-Fock (HF) single-determinant $\ket{\Phi_r}$. \cite{bartlett-RMP2007}
%
Ground state quantum Monte Carlo (QMC) stochastic methods,
\cite{Ceperley1980,Reynolds1982,Foulkes2001,SZ-HK:2003} which are exact
in principle, use projection from any $\ket{\Phi_r}$ that has non-zero overlap
with the ground state wave function (WF).
In practice, however, the Fermionic sign
problem \cite{Ceperley_sign,kalos91,Zhang1999_Nato,Foulkes2001,SZ-HK:2003}
must be controlled to achieve accurate results.
Diffusion QMC (DMC) uses a single- or multi-reference WF to
impose approximate Fermionic nodal boundary conditions in real
space and also includes a Jastrow factor to reduce the
stochastic variance.
The recently developed phaseless auxiliary-field quantum Monte Carlo
(AFQMC) method
\cite{SZ-HK:2003,Al-Saidi_TiO-MnO:2006,Al-Saidi_GAFQMC:2006,QMC-PW-Cherry:2007}
is an alternative and complementary QMC approach, which samples the
many-body wave function with random walkers in the space of Slater
determinants. AFQMC provides a different route to controlling the
sign problem, using the complex overlap of the walkers with $\ket{\Phi_r}$,
which is frequently just a single HF determinant.
Like the CC method, 
the AFQMC method works in a chosen 
single-particle basis,
and it has been successfully applied using Gaussian
\cite{Al-Saidi_GAFQMC:2006,Al-Saidi_Post-d:2006,Al-Saidi_H-bonded:2007,Al-Saidi_Bondbreak:2007}
and plane wave
\cite{SZ-HK:2003,Al-Saidi_TiO-MnO:2006,QMC-PW-Cherry:2007} basis sets.

While these correlated methods are generally quite accurate near
equilibrium geometries, the use of an approximate $\ket{\Phi_r}$ can introduce
uncontrolled errors 
as bonds are stretched or broken. 
\cite{DavidsonErnest_symbreak_1983,Umrigar-SpnContam-1998,sears_spncontam_F2_2003,lochanHeadGordon_spncontam_2007}
The main reason for this is that correlation effects become increasingly
important in the transition region where a system begins to dissociate
into its fragments, which are themselves often open shell systems. The
quality of $\ket{\Phi_r}$ typically degrades in this region, since it is derived
from a simple level of theory.
A second reason is the
breaking of spin or spatial symmetries in these simple reference WFs.

In previous applications, phaseless AFQMC with an unrestricted Hartree-Fock
(UHF) single-determinant $\ket{\Phi^{\textrm{UHF}}_r}$ was found to
often give better overall and more uniform accuracy than CCSD(T) in
mapping the potential-energy curve (PEC).
\cite{Al-Saidi_GAFQMC:2006,Al-Saidi_Post-d:2006,Al-Saidi_H-bonded:2007}
In some cases, however, such as the BH and N$_2$ molecules, achieving
quantitative accuracy of a few $\mEh$ for the entire PEC required
multi-determinant $\ket{\Phi_r}$.
\cite{Al-Saidi_Bondbreak:2007}
In these cases, spin contamination
did not appear to be a major source of the error seen in the calculations
with UHF reference states.
In this paper, we show that,
with a single-determinant $\ket{\Phi^{\textrm{UHF}}_r}$, the AFQMC potential
energy curve of the difficult \Fz{} molecule is
{qualitatively} incorrect 
in the intermediate dissociation region.
Spin contamination of $\ket{\Phi^{\textrm{UHF}}_r}$ is found to be the
dominant factor for this error.
We describe a simple spin-projection method to effectively remove spin-contamination effects.
\COMMENTED{
This is
comparable to or better than those obtained with a multi-determinant
$\ket{\Phi_r}$ trial WF from GVB or CASSCF.
The spin-projection method thus further reduces the reliance of AFQMC on the choice of
$\ket{\Phi_r}$, which is one of its most desirable features.
}

With $\ket{\Phi^{\textrm{UHF}}_r}$ and the spin-projection method,
the AFQMC results of \Fz{} are shown to be accurate (within a few $\mEh$ of the
near-exact CCSDTQ result)
across the entire PEC.
One of the main appeals of QMC methods is that
the computational cost typically scales with systems size as a low power.
Using larger basis sets (cc-pVTZ and cc-pVQZ), we then obtain
realistic PECs and spectroscopic constants
and compare them with experimental results.

The remainder of the paper is organized as
follows. Section~\ref{sec:spncontam-problem} discusses
the difficulties in calculating accurate \Fz{} PECs. In Section~\ref{sec:solved} a simple method
is described that removes spin-contamination effects in AFQMC calculations. Realistic
\Fz{} potential energy curves and spectroscopic constants are presented in
Section~\ref{sec:spectroscopic}.
Finally, Section~\ref{sec:summary} summarizes and discusses our
principal results.

\section{Spin contamination effects in the dissociation of the \Fz{} molecule}
\label{sec:spncontam-problem}

\begin{figure}[!htbp]
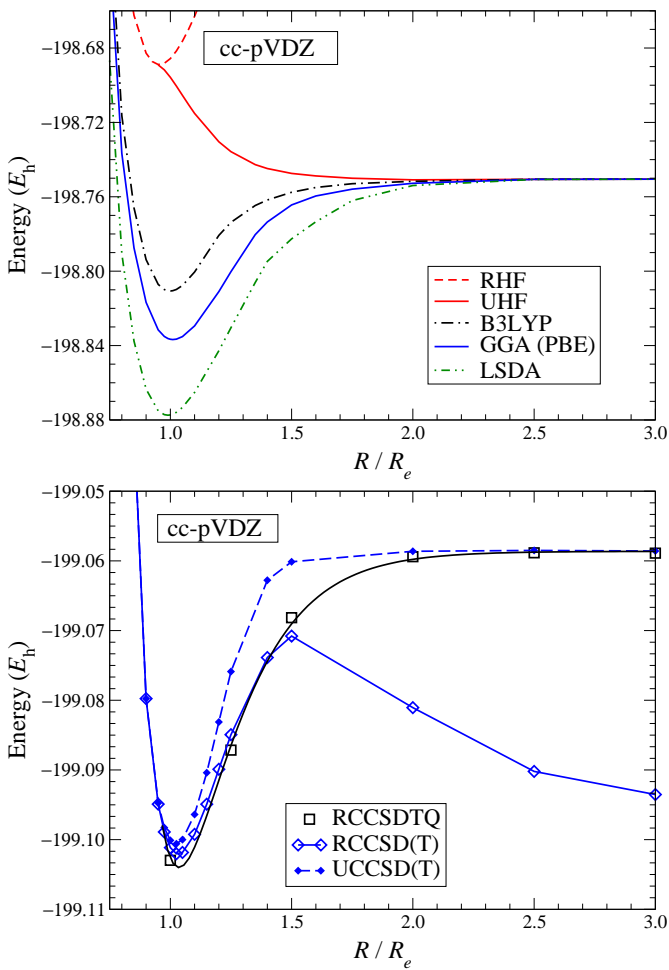

\includegraphics[scale=.35]{\FIGDIR{}F2-cc-pVDZ-meanfield.eps}
\\
\includegraphics[scale=.35]{\FIGDIR{}F2-cc-pVDZ-corr.eps}
\caption{(Color online) PECs for \Fz{} using the cc-pVDZ basis.
The F--F internuclear distance, $R$, is shown in units of the
equilibrium value $R_e \equiv 1.41193$\,\AA\,.\cite{Huber1979}
(Note the energy scales are different
between the upper and lower figures.)
Upper figure: mean-field results, with constant shifts added to the DFT
energies so that all energies match the UHF energy at $R / R_e = 3.0$.
Lower figure: coupled-cluster results, both spin-restricted [RCCSD(T)
and RCCSDTQ] and spin-unrestricted [UCCSD(T)]. 
The RCCSDTQ results
are from Ref.~\onlinecite{Musial2005}. A Morse fit through the
RCCSDTQ points is shown as a guide to the eye. Straight line segments
connect the RCCSD(T) and UCCSD(T) points.
}
\label{fig:F2-cc-pVDZ-stdQC}
\end{figure}

The difficulty in treating the dissociation of the \Fz{} molecule is
already evident at the mean-field level of theory.
The upper panel of Fig.~\ref{fig:F2-cc-pVDZ-stdQC} shows PECs from HF and
density functional theory (DFT).
UHF does not predict a bound molecule,\cite{Hijikata1961,Gordon1987}
while the restricted HF (RHF) curve is artificially bound
with a minimum that is $5\%$ too low compared to experiment.
The DFT local spin-density approximation \cite{PerdewLocal} (LSDA)
and generalized gradient approximation \cite{Perdew1996} (GGA/PBE)
yield dissociation energies which are too large.
The hybrid B3LYP\cite{Becke1993, Stephens1994} dissociation energy
is closer to experiment, but the shape of
the B3LYP PEC is not correct in the intermediate region (see below
in Sec.~\ref{sec:spectroscopic}).
All our HF and DFT calculations were carried out using the
quantum chemistry computer program \GAUSSIAN[98].\cite{Gau98}

The difficulty of treating \Fz{} dissociation, even using correlated
methods, is illustrated in the bottom panel of
Fig.~\ref{fig:F2-cc-pVDZ-stdQC}. 
For the small (cc-pVDZ)
basis set \cite{Dunning1989, EMSL_BasisSets2007} 
chosen here, the spin-restricted coupled-cluster including up to
quadrupole excitations (RCCSDTQ) is within reach, which
is expected to give close to exact results in this case. 
The RCCSD(T) and UCCSD(T) calculations,
done with \GAUSSIAN[98] \cite{Gau98} or {\NWCHEM}, \cite{NWChem-4.6}
use a single-determinant RHF or UHF $\ket{\Phi_r}$, respectively.
The RCCSD(T) method breaks down in the dissociation limit,
where the RHF $\ket{\Phi_r}$ is very poor (as seen in the upper panel of the
figure). The UCCSD(T) PEC is accurate in the dissociation limit, but
its shape begins to be distorted
near the equilibrium bond length
and shows significant error in the intermediate region. 

\begin{figure}[!htbp]
\includegraphics[scale=.35]{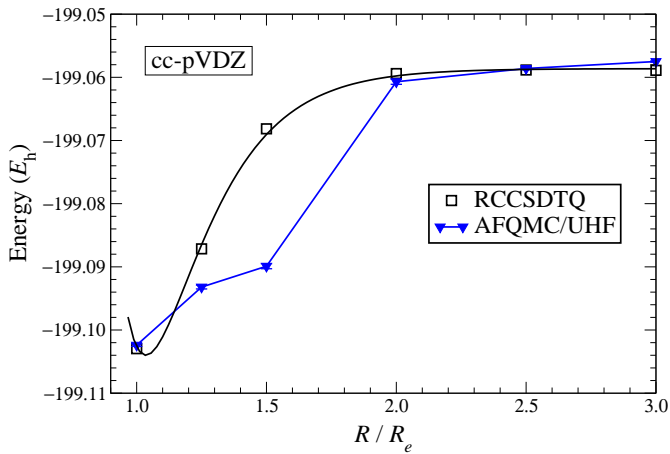}
\caption{(Color online) AFQMC \Fz{} PEC using a UHF reference
state WF, compared to RCCSDTQ results from
Fig.~\ref{fig:F2-cc-pVDZ-stdQC}.
The cc-pVDZ basis set is used. A Morse fit through the RCCSDTQ
points is shown as a guide to the eye. Straight line segments connect
the AFQMC/UHF points.  The large deviations 
at $R / R_e \sim 1.5$ are due to spin contamination of the UHF reference
state WF (see text).}
\label{fig:F2-QMC-spin-contm}
\end{figure}

The AFQMC PEC calculated with
$\ket{\Phi^{\textrm{UHF}}_r}$ (labeled AFQMC/UHF) shows good agreement
with RCCSDTQ near equilibrium and in the dissociation limit, as seen in
Fig.~\ref{fig:F2-QMC-spin-contm}.
In the intermediate regime, however, 
AFQMC/UHF shows deviations of more than $20\,\mEh$.  The poor results
in this regime are due to the AFQMC phase-free approximation \cite{SZ-HK:2003}
when it is applied to a walker population that is spin-contaminated.
The approximation, which depends on the accuracy of $\ket{\Phi_r}$, is analogous
to that in the fixed node DMC method, whose performance depends on
the accuracy of the $\ket{\Phi_r}$ nodal hypersurface.  In view of the inability of
$\ket{\Phi^{\textrm{UHF}}_r}$
to even bind F$_2$ and its poor quality in the intermediate
regime, the inaccurate AFQMC results are perhaps not too surprising.

\begin{figure}[!hbtp]
\includegraphics[scale=.35]{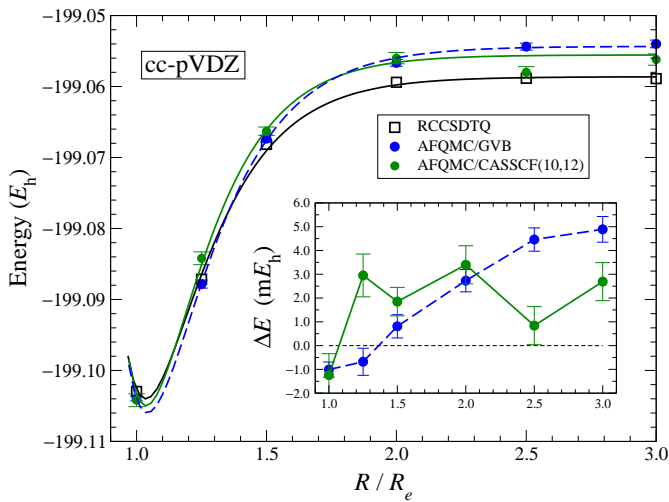}
\caption{(Color online) Improvement of the AFQMC \Fz{} PEC using
multi-determinant reference state WFs from GVB(1/2) and CASSCF(10,12).
RCCSDTQ from
Fig.~\ref{fig:F2-cc-pVDZ-stdQC} is shown for comparison.  Morse fits are shown as a guide to
the eye. The inset shows energy differences $\Delta E$ (in $\mEh$)
compared to RCCSDTQ. All calculations use the cc-pVDZ basis set.}
\label{fig:F2-cc-pVDZ-QMC-psiT-try}
\end{figure}

A brute force approach to improve the AFQMC PEC is to use a better
$\ket{\Phi_r}$, through the use of a multi-determinant reference wave function.
Indeed, using a generalized valence bond (GVB)\cite{Bobrowicz1977}
or complete active space self consistent field (CASSCF)\cite{Schmidt1998}
$\ket{\Phi_r}$ in AFQMC (labeled AFQMC/GVB and AFQMC/CASSCF,
respectively) eliminates most of the error, as shown in
Fig.~\ref{fig:F2-cc-pVDZ-QMC-psiT-try}.
The GVB WF is a perfect-pairing GVB(1/2) wave function, where the
electron pair responsible for the chemical bonding in \Fz{} (those in the
$2p_z\,\sigma_g$ orbital in RHF) now occupy a pair of nonorthogonal,
$2p_z$-atomic-like orbitals. The GVB WF
has the proper dissociation limit. 
%
The CASSCF(10,12) is for 10 active electrons and an active space of 12
molecular orbitals.
The CASSCF WF is truncated,
retaining only those determinants whose
weights (the square of the configuration-interaction coefficients)
are greater than $4 \times 10^{-4}$.
(The adequacy of this cutoff was tested by performing additional
calculations including determinants with weights larger than $10^{-4}$ in
the trial WF.
Within statistical errors, QMC energies similar to those with the higher
weight cutoff were obtained.)
The computational cost in AFQMC with a multi-determinant $\ket{\Phi_r}$
scales linearly with the number of determinants, although
the real cost is typically less since
a better $\ket{\Phi_r}$ reduces statistical errors.
\cite{Al-Saidi_Bondbreak:2007}
In the next section, we show that the improved AFQMC/GVB and
AFQMC/CASSCF PECs are largely due to the elimination of errors from
spin contamination.

\COMMENTED{ 
In AFQMC/UHF, the UHF trial WF is spin-contaminated.  The UHF
expectation value of the total electronic spin operator $\Sop^2$ is
$0.362$, $0.978$, and $1.004$ at $R/Re = 1.0$, $1.5$, and $3.0$,
respectively, indicating that the triplet component grows as the
molecule is stretched.  The GVB $\ket{\Phi_r}$ is spin-contamination
free by design, and the truncated CASSCF $\ket{\Phi_r}$ is nearly free
of spin contamination.

}

\section{Eliminating spin contamination in AFQMC: spin projection method}
\label{sec:solved}

\begin{table*}[!htbp]
\caption{\label{tbl:F2-benchmark}
Comparison of computed \Fz{} PEC for various methods, using the cc-pVDZ basis.
The RCCSDT and RCCSDTQ results are from Ref.~\onlinecite{Musial2005}.
Energies are in $E_{\textrm{h}}$. QMC statistical errors are on the last
digit and are shown in parentheses.
}

\begin{tabular}{l r@{}l r@{}l r@{}l r@{}l r@{}l r@{}l}
\hline
\hline
 & \multicolumn{12}{c}{$R / R_e$} \\
\cline{2-13}
 & \multicolumn{2}{c}{$1.0$}
 & \multicolumn{2}{c}{$1.25$}
 & \multicolumn{2}{c}{$1.5$}
 & \multicolumn{2}{c}{$2.0$}
 & \multicolumn{2}{c}{$2.5$}
 & \multicolumn{2}{c}{$3.0$} \\
\hline
%
RHF                   & $-198$&$.685670$  & $-198$&$.612171$  & $-198$&$.527711$  & $-198$&$.419839$  & $-198$&$.374025$  & $-198$&$.355748$  \\
UHF                   & $-198$&$.695746$  & $-198$&$.735754$  & $-198$&$.747441$  & $-198$&$.750892$  & $-198$&$.750597$  & $-198$&$.750518$  \\
GVB                   & $-198$&$.761466$  & $-198$&$.759320$  & $-198$&$.748618$  & $-198$&$.743801$  & $-198$&$.743570$  & $-198$&$.743650$  \\
CASSCF(10,12)         & $-198$&$.886738$  & $-198$&$.874892$  & $-198$&$.857896$  & $-198$&$.850284$  & $-198$&$.849831$  & $-198$&$.849732$  \\
RCCSD(T)              & $-199$&$.101152$  & $-199$&$.084940$  & $-199$&$.070790$  & $-199$&$.081058$  & $-199$&$.090213$  & $-199$&$.093534$  \\
UCCSD(T)              & $-199$&$.100100$  & $-199$&$.075878$  & $-199$&$.060126$  & $-199$&$.059302$  & $-199$&$.058784$  & $-199$&$.058687$  \\
RCCSDT                & $-199$&$.101417$  & $-199$&$.084493$  & $-199$&$.065170$  & $-199$&$.057558$  & $-199$&$.058023$  & $-199$&$.057933$  \\
RCCSDTQ               & $-199$&$.102961$  & $-199$&$.087149$  & $-199$&$.068153$  & $-199$&$.0594$    & $-199$&$.05884$   & $-199$&$.05889$   \\
AFQMC/UHF             & $-199$&$.1024(2)$ & $-199$&$.0932(3)$ & $-199$&$.0899(4)$ & $-199$&$.0607(4)$ & $-199$&$.0586(2)$ & $-199$&$.0575(1)$ \\
AFQMC/GVB             & $-199$&$.1040(3)$ & $-199$&$.0878(6)$ & $-199$&$.0673(5)$ & $-199$&$.0567(5)$ & $-199$&$.0544(5)$ & $-199$&$.0540(5)$ \\
AFQMC/CASSCF(10,12)   & $-199$&$.1042(9)$ & $-199$&$.0842(9)$ & $-199$&$.0663(6)$ & $-199$&$.0560(8)$ & $-199$&$.0580(8)$ & $-199$&$.0562(8)$ \\
sp-AFQMC/UHF          & $-199$&$.1020(5)$ & $-199$&$.0876(7)$ & $-199$&$.0686(7)$ & $-199$&$.0574(2)$ & $-199$&$.0558(3)$ & $-199$&$.0562(5)$ \\
%
%
%
%
\hline
\hline
\end{tabular}
\end{table*}

While the exact eigenstates of a spin-independent non-relativistic
electronic Hamiltonian are eigenstates of the total spin operator
$\Sop^2$ and its $z$ component $\Sop_z$, approximate wave functions
may not be eigenstates of $\Sop^2$
unless special care is taken.
Such
approximate wave functions are called \textit{spin contaminated}.  For
simplicity, we restrict the discussion in this section to the case
where the reference state wave function is given by
$\ket{\Phi^{\textrm{UHF}}_r}$ with $S_z = 0$,
approximating the exact singlet ground state $\ket{\Phi^s_0}$. In this case $\ket{\Phi^{\textrm{UHF}}_r}$
will generally be spin contaminated, \textit{i.e.}, containing triplet $\ket{\Psi^t}$ and higher spin states:
\eql{eq:UHF-singlet}
{
     \ket{\Phi^{\textrm{UHF}}_r}  = c_s \ket{\Psi^s} + c_t \ket{\Psi^t} + \ldots
  \,,
}
where $\ket{\Psi^s}$ is a linear combination of the ground and excited singlet states.
In the UHF result of
Fig.~\ref{fig:F2-cc-pVDZ-stdQC}, the
expectation value of the total electronic spin operator $\Sop^2$
in $\ket{\Phi^{\textrm{UHF}}_r}$ is $0.362$, $0.978$, and $1.004$ at
$R/R_e = 1.0$, $1.5$, and $3.0$, respectively, indicating a high level of
spin contamination in which the triplet
component grows as the molecule is stretched.

Ideally, AFQMC projection of $\ket{\Phi^{\textrm{UHF}}_r}$ would lead to the exact spin-contamination-free
ground state, since
\eql{eq:QMC-proj-spincontm} { \left(e^{-\tau\Hop}\right)^n
\ket{\Phi^{\textrm{UHF}}_r} \to C_0 \ket{\Phi^s_0} + C_1 e^{- n\tau (E_1-E^s_0)} \ket{\Phi_1} \, ,
}
where
$\ket{\Phi_1}$ is the exact first excited state, $\tau$ is the time-step
parameter, and as $n \to \infty$, all
components except $\ket{\Phi^s_0}$ become vanishingly small. The use
of the phase-free approximation, \cite{SZ-HK:2003} however, effectively
modifies this projection so that a triplet component can
survive. Thus, the population of AFQMC random walkers will be
spin contaminated if it was initialized with 
$\ket{\Phi^{\textrm{UHF}}_r}$. In F$_2$, the presence of a
nearby triplet state \cite{Cartwright1979} at bond lengths $R/R_e \gtrsim 1.4$
exacerbates this, and this is where the AFQMC/UHF PEC shows the largest
error.

In the previous section, spin contamination in AFQMC was
eliminated through the use of a (nearly) spin-pure multi-determinant $\ket{\Phi_r}$, which effectively
filters the population of random walkers, retaining only the spin-pure
component regardless of how the population was initialized.
The GVB $\ket{\Phi_r}$ is spin-contamination free by design,
and the truncated CASSCF $\ket{\Phi_r}$ is nearly free of spin contamination.
The elimination/reduction of
spin contamination in the GVB and truncated CASSCF $\ket{\Phi_r}$
is a main factor in the improvement of the corresponding QMC results.
A strong clue to this is seen in the case with GVB $\ket{\Phi_r}$,
where the GVB WF has only two determinants and has a variational
energy within $\sim 1\,\mEh$ that of the UHF at $R/R_e = 1.5$
(see Table~\ref{tbl:F2-benchmark}), and yet
QMC/GVB greatly improves over QMC/UHF.

\begin{figure}[!tbp]
\includegraphics[scale=.35]{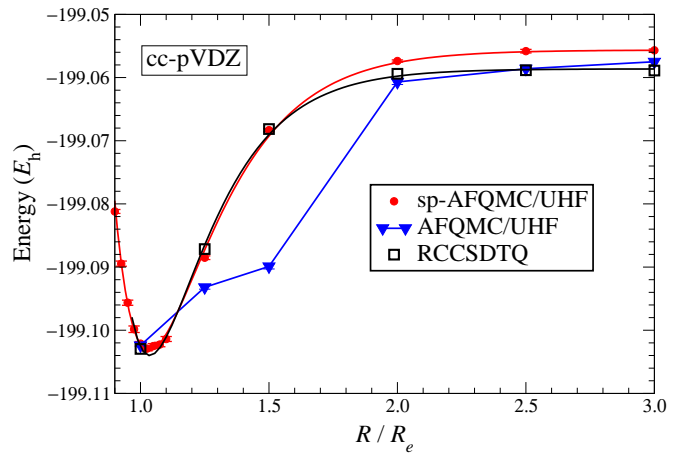}
\caption{(Color online) Improvement of the AFQMC \Fz{} PEC using
spin projection with a single determinant UHF reference state wave
function. Spin-projected (sp-AFQMC/UHF) results are compared to standard
AFQMC/UHF without spin projection and to RCCSDTQ results from Fig.~\ref{fig:F2-QMC-spin-contm}.
Morse fits are shown as a guide to the eye, except for AFQMC/UHF. All
calculations use the cc-pVDZ basis.}
\label{fig:F2-cc-pVDZ-QMC}
\end{figure}

A simpler way to eliminate spin contamination in AFQMC is to ensure
that the population of random
walkers consists of spin-pure (\textit{i.e.} RHF-type) Slater
determinants, $\{ \, \ket{\phi_s} \, \}$.
Almost all phaseless AFQMC electronic structure
calculations to date
\cite{SZ-HK:2003,Al-Saidi_GAFQMC:2006,QMC-PW-Cherry:2007,Al-Saidi_H-bonded:2007}
have used Hubbard-Stratonovich (HS) transformations which preserve spin
symmetry,\cite{note_BrokenSym_HS}
%
\begin{equation}
    \vop_{\textrm{HS}}(\mathbf{x})
  = \vop_\uparrow(\mathbf{x}) + \vop_\downarrow(\mathbf{x})
\,,
\label{eq:vHS}
\end{equation}
%
where $\mathbf{x}$ denotes HS auxiliary fields, and
the one-body operators $\vop_\uparrow(\mathbf{x})$ and
$\vop_\downarrow(\mathbf{x})$ have identical forms.
For example,
in the plane-wave formalism,\cite{SZ-HK:2003,QMC-PW-Cherry:2007}
$\vop_{\textrm{HS}}(\mathbf{x})$ is essentially a Fourier component of
the density operator.
Thus, if a random walker is initialized to a spin-pure Slater determinant
with $S_z = 0$,
its spin state
cannot be modified by the QMC propagation $\ket{\phi'_s} =
e^{-\vop_{\textrm{HS}}(\mathbf{x})}\ket{\phi_s} $.
Typically, the same trial WF is used in phaseless AFQMC to generate the initial
population, to guide the importance sampling, and to impose the
phaseless constraint. \cite{SZ-HK:2003,sz-cpmc}
This of course does not have to be the case.
Here we use a spin-pure state to initialize the walkers. 
Since each walker in the
population $\{ \, \ket{\phi_s} \, \}$ remains spin-pure, the local
energy $E_L[\phi_s]$ projects out the 
triplet and higher components
of $\ket{\Phi^{\textrm{UHF}}_r}$:
\begin{equation}
 E_L[\phi_s] = \frac{\langle \Phi^{\textrm{UHF}}_r |\hat{H}|\phi_s\rangle}
                    {\langle \Phi^{\textrm{UHF}}_r  | \phi_s\rangle} \,.
\label{eq:El}
\end{equation}
The mixed estimator for the ground state energy is determined by the
local energy, so it too is spin-uncontaminated. 
Thus, higher spin states
have no effect on either the AFQMC projection, the phase-free
approximation, or the ground state energy estimation.

The spin-projected AFQMC (sp-AFQMC) method
described above shows a dramatic improvement over the spin-contaminated
AFQMC/UHF in \Fz{}, as seen in Fig.~\ref{fig:F2-cc-pVDZ-QMC}.
In the sp-AFQMC/UHF calculations, the walker population is
initialized with the RHF solution $\ket{\phi_s} = \ket{\textrm{RHF}}$,
but $\ket{\Phi^{\textrm{UHF}}_r}$ is used to implement the phase-free
constraint \cite{SZ-HK:2003} and to calculate the local 
energy.
Table~\ref{tbl:F2-benchmark} tabulates the energies for all methods
using the cc-pVDZ basis.
We see that the sp-AFQMC/UHF PEC is in excellent agreement with the
RCCSDTQ result, with a maximum discrepancy of $\sim 3\,\mEh$. This
accuracy is in fact slightly better than that of
either AFQMC/CASSCF or AFQMC/GVB.

\COMMENTED{
To test the spin-projected AFQMC method described above, we first
applied it to \Fz{}, using the cc-pVDZ basis. The walker population is
initialized with the RHF solution $\ket{\phi_s} = \ket{\textrm{RHF}}$,
but $\ket{\Phi^{\textrm{UHF}}_r}$ is used to implement the phase-free
constraint \cite{SZ-HK:2003} and to calculate the local 
energy.  We choose $\ket{\Phi^{\textrm{UHF}}_r}$ as the reference state,
because the phase-free approximation delivers the most accurate
results when the best available variational $\ket{\Phi_r}$ is used, as found
previously. \cite{Al-Saidi_GAFQMC:2006}
For a single-determinant $\ket{\Phi_r}$, this is the UHF solution. As
seen in Fig.~\ref{fig:F2-cc-pVDZ-QMC}, the spin-projected results
(sp-AFQMC/UHF) show a dramatic improvement over the original AFQMC/UHF
PEC, with accuracy comparable to that shown in Fig.~\ref{fig:F2-cc-pVDZ-QMC-psiT-try}
from a multi-determinant $\ket{\Phi_r}$.
Table~\ref{tbl:F2-benchmark} tabulates the energies for all methods
using the cc-pVDZ basis.
}

\begin{figure}[!tbp]
\includegraphics[scale=.35]{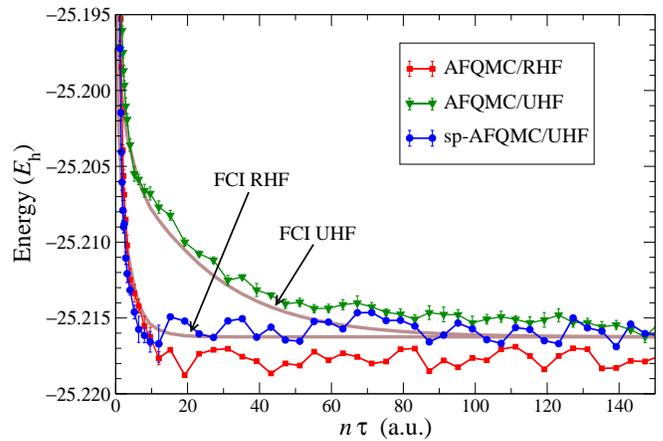}
\caption{(Color online) Slow energy equilibration as a function of
imaginary time $n \tau$ of
spin contaminated AFQMC/UHF, compared to spin contamination free
AFQMC/RHF and sp-AFQMC/UHF for the BH molecule at $R_{e}=1.2344$\,\AA.
The cc-pVDZ basis is used.
To reduce clutter, QMC statistical errors are not shown in
sp-AFQMC/UHF and AFQMC/RHF for $n \tau>15$, but the average
size of the error bar is indicated for each in the legend.
FCI-derived RHF and UHF projection curves (see text) are also shown,
calculated using Eq.~(\ref{eq:QMC-proj-spincontm}).
The FCI ground-state energy is $-21.216\,249\,{E_\textrm{h}}$.
}

\label{fig:betaequilibration}
\end{figure}

In addition to removing spurious spin contamination effects in the
calculated AFQMC energy, sp-AFQMC can sometimes also reduce the
imaginary time [see Eq.~(\ref{eq:QMC-proj-spincontm})] needed to
obtain energy equilibration.  This is illustrated in
Fig.~\ref{fig:betaequilibration} for the BH molecule.
The spin-contaminated AFQMC/UHF has an equilibration time \mbox{$n \tau \sim 100$\,a.u.},
about an order of magnitude larger than the
spin-contamination-free sp-AFQMC/UHF (\mbox{$n \tau \sim 10$\,a.u.}).
For comparison, the curve from AFQMC using
$\ket{\Phi^{\textrm{RHF}}_r}$ is also shown.
Starting from the same initial state, the the spin-contamination-free AFQMC/RHF has a short
equilibration time similar to sp-AFQMC/UHF, but the
converged result has a larger systematic error, because of the poorer
quality of $\ket{\Phi^{\textrm{RHF}}_r}$ as the constraining WF in the phaseless approximation.
The different behaviors of
the equilibration time can be understood 
by comparing with FCI-derived RHF and UHF projections, which are shown
in Fig.~\ref{fig:betaequilibration}.
We calculate the ``exact'' projection results by
expanding the UHF and RHF initial WFs in terms of a truncated
set of the FCI eigenstates
(the first 80 eigenstates, obtained with \GAMESS\cite{Gamess}).
With a UHF initial WF, the long equilibration time is
due to the presence of
low-lying triplet
components 
[see Eq.~(\ref{eq:UHF-singlet})],
which results in smaller effective gap
$(E_1-E^s_0)$ in Eq.~(\ref{eq:QMC-proj-spincontm}).
The RHF WF, on the other hand, has no overlap with any
triplet state, and consequently the effective gap is larger
and the equilibration time shorter.
\COMMENTED{
We have found similar results for other small molecules.
Thus, calculations using spin-projected AFQMC are also more efficient, requiring far
fewer imaginary time steps $n \tau$ in the initial equilibration of the random walker population.
}

\begin{table}[!htbp]
\caption{\label{tbl:F2-cc-pVQZ-PES}
The PEC of \Fz{}, computed using spin-projected phaseless AFQMC with UHF trial wave
function in cc-pVTZ and cc-pVQZ basis.
Energies are in $E_{\textrm{h}}$.
QMC statistical errors are on the last digit and are shown in parentheses.
}
\begin{tabular}{l r@{}l c r@{}l}
\hline
\hline
$R / R_e$   &
 \multicolumn{2}{c}{cc-pVTZ} &&
 \multicolumn{2}{c}{cc-pVQZ} \\
\hline
$0.80$      & $-199$&$.2321(4)$ && $-199$&$.3372(5)$ \\
$0.85$      & $-199$&$.2805(5)$ && $-199$&$.3824(6)$ \\
$0.9$       & $-199$&$.3061(6)$ && $-199$&$.4099(3)$ \\
$0.925$     & $-199$&$.3134(4)$ && $-199$&$.4159(6)$ \\
$0.95$      & $-199$&$.3186(5)$ && $-199$&$.4218(7)$ \\
$0.975$     & $-199$&$.3210(6)$ && $-199$&$.4231(5)$ \\
$1.0$       & $-199$&$.3211(4)$ && $-199$&$.4241(4)$ \\
$1.025$     & $-199$&$.3209(4)$ && $-199$&$.4236(4)$ \\
$1.05$      & $-199$&$.3190(4)$ && $-199$&$.4219(4)$ \\
$1.075$     & $-199$&$.3176(5)$ && $-199$&$.4201(4)$ \\
$1.1$       & $-199$&$.3136(5)$ && $-199$&$.4170(4)$ \\
$1.25$      & $-199$&$.2962(6)$ && $-199$&$.3975(4)$ \\
$1.4$       & $-199$&$.2797(6)$ && $-199$&$.3807(3)$ \\
$1.5$       & $-199$&$.2731(6)$ && $-199$&$.3732(4)$ \\
$1.75$      & $-199$&$.2632(4)$ && $-199$&$.3641(2)$ \\
$2.0$       & $-199$&$.2603(6)$ && $-199$&$.3605(4)$ \\
$3.0$       & $-199$&$.2590(6)$ && $-199$&$.3593(4)$ \\
\hline
\hline
\end{tabular}
\end{table}

\begin{figure*}[!hbtp]
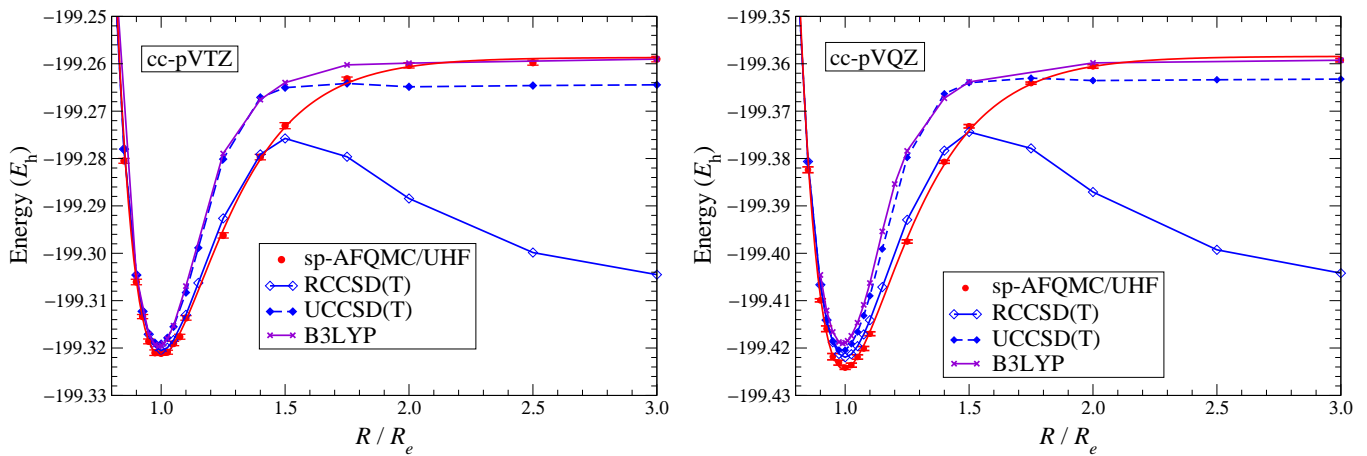

\includegraphics[scale=.35]{\FIGDIR{}F2-cc-pVTZ-spQMC-UHF.eps}\quad
\includegraphics[scale=.35]{\FIGDIR{}F2-cc-pVQZ-spQMC-UHF.eps}
\caption{(Color online) PECs for \Fz{} with cc-pVTZ and cc-pVQZ basis sets.
A Morse fit passes through the sp-AFQMC/UHF points.
Straight line
segments connect results of the other methods. The B3LYP curves were
shifted to agree with sp-AFQMC/UHF at $R/R_e = 3$.
%
%
}
\label{fig:F2-cc-pVQZ-QMC}
\end{figure*}

\begin{table*}[!htbp]
\caption{\label{tbl:F2-consts}
Computed \Fz{} spectroscopic constants for three basis sets, together
with experimental results.
}
\begin{tabular}{l lllllll}
\hline
\hline
%
 & {Expt\FN[1]}
 & {AFQMC}
 & {RCCSD(T)}
 & {UCCSD(T)}
 & {LSDA}
 & {GGA/PBE}
 & {B3LYP} \\
\hline
\multicolumn{6}{l}
{Basis: cc-pVDZ} \\
$r_e$ (\AA)           & 1.4131(8)&    1.467(4)  &    1.4571 &    1.4428 &    1.3970 &    1.4276 &    1.4097 \\
$\omega_0$ (cm$^{-1}$)& 916.64  &   725(36)     &    785    &  853      & 1026      &  958      & 1033      \\
$D_e$ (eV)\FN[2]      & 1.693(5)&     1.293(7)  &1.180\FN[3]&    1.142  &    3.454  &    2.347  &    1.638  \\
%
\hline
\multicolumn{6}{l}
{Basis: cc-pVTZ} \\
$r_e$ (\AA)           & 1.4131(8)&    1.411(3)  &    1.4131 &    1.3987 &    1.3863 &    1.4138 &    1.3957 \\
$\omega_0$ (cm$^{-1}$)& 916.64  &   928(30)     &  926      & 1022      & 1065      & 1001      & 1072      \\
$D_e$ (eV)\FN[2]      & 1.693(5)&     1.70(2)   &    --     &    1.493  &    3.486  &    2.351  &    1.651  \\
$D_e$ (eV)\FN[3]      & 1.693(5)&     1.60(1)   &    1.523  &    1.495  &           &           &           \\
\hline
\multicolumn{6}{l}
{Basis: cc-pVQZ} \\
$r_e$ (\AA)           & 1.4131(8)&    1.411(2)  &    1.4108 &    1.3946 &    1.3856 &    1.4136 &    1.3944 \\
$\omega_0$ (cm$^{-1}$)& 916.64  &   912(11)     &  929      & 1036      & 1062      &  997      & 1109      \\
$D_e$ (eV)\FN[2]      & 1.693(5)&     1.77(1)   &    --     &    1.567  &    3.473  &    2.321  &    1.634  \\
$D_e$ (eV)\FN[3]      & 1.693(5)&     1.70(1)   &    1.594  &    1.569  &           &           &           \\
\hline
\hline
\footnotetext[1]{
    The dissociation energy $D_e$ is from Ref.~\onlinecite{Bytautas2005}
    (the zero point and spin-orbit energies have been removed). The
    equilibrium internuclear distance $r_e$ is from Ref.~\onlinecite{Edwards1976}, and the
    vibrational frequency $\omega_0$ is from Ref.~\onlinecite{Huber1979}.
    \REMARKS{Still need to find errorbar for $\omega_0$ in literature.
    The $\omega_0$ from Huber\&Herzberg is obtained from fitting and
    carries no errorbar.}
}
\footnotetext[2]{
    The dissociation energy calculated using $E(3R_e) - E(r_e)$.
}
\footnotetext[3]{
    The dissociation energy calculated using $2 E(\textrm{atom}) - E(r_e)$.
    In AFQMC, $E(\textrm{atom})$ is calculated with
    a truncated CASSCF(7,13) $\ket{\Phi_r}$.
}
\end{tabular}
\end{table*}

\section{Realistic \Fz{} potential energy curve: Basis-set converged spin-projected AFQMC results}
\label{sec:spectroscopic}

We have shown that the sp-AFQMC
PEC is accurate at the double zeta cc-pVDZ level, where
near-exact CCSDTQ coupled-cluster results are available for comparison.
As a function of bond stretching, sp-AFQMC delivers more uniform accuracy than RCCSD(T)
and UCCSD(T) for the difficult F$_2$ molecule, with absolute errors of
a few $\mEh$ or less.
In this section, we employ large basis sets to obtain a realistic PEC.
We also compute
F$_2$ spectroscopic constants and compare them with experimental results.

Figure~\ref{fig:F2-cc-pVQZ-QMC} presents the PECs of \Fz{} computed
using sp-AFQMC/UHF for cc-pVTZ and cc-pVQZ basis
sets. \cite{EMSL_BasisSets2007} For comparison, PECs from B3LYP,
RCCSD(T) and UCCSD(T) are also shown, representing the best current
theoretical results. (The B3LYP curves
were shifted to agree with sp-AFQMC/UHF at $R/R_e = 3$.)
The sp-AFQMC/UHF energies corresponding to
Fig.~\ref{fig:F2-cc-pVQZ-QMC} are also tabulated in
Table~\ref{tbl:F2-cc-pVQZ-PES}.

Computed spectroscopic constants are given in
Table~\ref{tbl:F2-consts} together with those from the many-body RCCSD(T) and
UCCSD(T), and the independent-electron
LSDA, GGA/PBE, and B3LYP methods.
The spectroscopic constants were obtained by fitting
the calculated PECs in the range $0.8 \le R / R_e \le 1.25$
to a three-term extended Morse curve \cite{Coolidge1938}
\eql{eq:ext4Morse}
{
    E(r) = E_0 +
    \sum_{n=2}^{4}
    \frac{C_n}{a^n}
    \left[
      1 - e^{-a(r - r_e)}
    \right]^n.
}
The fitting procedure yields the molecular electronic energy,
$E_0 \equiv E(r_e)$, equilibrium bond length $r_e$, and the
harmonic frequency $\omega_0 = \sqrt{C_2 / 2 \mu}$, where
$\mu$ is the reduced mass of the {\Fz} molecule.
The dissociation energy is given by $D_e \equiv E(3R_e)-E(r_e)$.
For comparison, $D_e$ calculated from
$2 E(\textrm{atom}) - E(r_e)$, where $E(\textrm{atom})$ is a well-converged
energy for the isolated atom, is also shown
for the many-body results in the TZ and QZ basis sets.

The values of sp-AFQMC/UHF $r_e$ and $\omega_0$
in Table~\ref{tbl:F2-consts} are in excellent agreement with experiment.
The dissociation energy \mbox{$D_e = E(3R_e)-E(r_e)$} is overestimated,
however.
This is due to the overestimation of the total energy at large $R / R_e = 3$,
which reflects the deficiency of a simple UHF $\ket{\Phi_r}$ in AFQMC for
open-shell systems, as previously noted.\cite{QMC-PW-Cherry:2007}
To obtain a more accurate $\De$, an AFQMC calculation was performed for
the isolated F atom with a truncated CASSCF(7,13) $\ket{\Phi_r}$.
The $2s$ through $3d$ orbitals were included in the active space of the
CASSCF WF. 
The truncation retains determinants with weight greater than $2 \times 10^{-4}$,
resulting in a $\ket{\Phi_r}$ with 47 determinants.
In cc-pVQZ, the atomic energy thus calculated is
\mbox{$E(\textrm{atom})= -99.6811(5)$\,\ensuremath{E_\textrm{h}}}, while the
corresponding RCCSD(T) and UCCSD(T) values are \mbox{$-99.681\,704$} and
\mbox{$-99.681\,576$\,\ensuremath{E_\textrm{h}}}, respectively.
The dissociation energy obtained with
\mbox{$D_e = 2 E(\textrm{atom}) - E(r_e)$}
is in excellent agreement with experiment.

The variations in the results from the TZ to the QZ basis sets
are still visible but quite small (especially in $r_e$ and $\omega_0$).
It is thus reasonable to expect the residual finite basis set error to be
small in the QZ basis.
A simple extrapolation to the infinite basis limit\cite{Inf-basis} increases $D_e$
only by $0.02$ eV ($0.7\,\mEh$) from the cc-pVQZ value.
The shape of the sp-AFQMC PEC should thus be
very close to that at basis set convergence.
(In contrast, the residual error of the cc-pVQZ absolute molecular
energies is approximately $110\,\mEh$, estimated using nonrelativistic
energies published in the literature.\cite{Filippi1996,Bytautas2005})

Compared to the sp-AFQMC/UHF PEC, the RCCSD(T) and UCCSD(T) PECs in
Fig.~\ref{fig:F2-cc-pVQZ-QMC} show the same shortcomings as seen with
the cc-pVDZ basis in Fig.~\ref{fig:F2-cc-pVDZ-stdQC}. While the
RCCSD(T) PEC near equilibrium is in good agreement with sp-AFQMC/UHF,
it is very poor in the dissociation limit.
For this reason, the RCCSD(T) dissociation energy $D_e$ shown in
Table~\ref{tbl:F2-consts} is computed only from
\mbox{$D_e = 2 E(\textrm{atom}) - E(r_e)$}.
The UCCSD(T) PEC is accurate in the dissociation limit, but its shape is
significantly distorted near equilibrium.
Consequently, the UCCSD(T) spectroscopic constants are not in as good
agreement with experiment.
The RCCSD(T) $r_e$ and $\omega_0$ are also in excellent agreement with
experiment, while the UCCSD(T) $\omega_0$ is $13\%$ too large.
This is consistent with
Fig.~\ref{fig:F2-cc-pVDZ-stdQC}, where the UCCSD(T) potential well is too
narrow compared with the near-exact RCCSDTQ result.
Curiously, the UCCSD(T) and B3LYP PECs show very similar deviations near
equilibrium.

\COMMENTED{
Both RCCSD(T) and UCCSD(T) underestimate $D_e$. These trends are
consistent with the comparison with the near-exact RCCSDTQ
results in Table~\ref{tbl:F2-benchmark} and
Fig.~\ref{sec:spncontam-problem}. and are most likely not
finite basis set errors.
and this is probably not
a finite basis set error, based on comparison with the near-exact RCCSDTQ
results in Table~\ref{tbl:F2-benchmark} and in
Fig.~\ref{sec:spncontam-problem}. Thus,
For example, RCCSD(T) and UCCSD(T) are
$\sim 2$ and $\sim 3$ $\mEh$ higher in energy at $R = R_e$, respectively, than
RCCSDTQ, and
the UCCSD(T) potential well is narrower (i.e., larger frequency $\omega_0$).
}

As expected, LSDA, GGA/PBE, and B3LYP show more rapid
convergence with basis set size than the correlated
methods.
The B3LYP $D_e$ is good, but since the shape of its PEC is incorrect,
$\omega_0$ is $\sim 20\%$ too large and the equilibrium bond length is too small.
(The large discrepancy here underscores the difficult nature of
\Fz{}; in other molecules,  B3LYP results
are typically found to be in good agreement with experiment.\cite{Sinnokrot2001})
Both LSDA and GGA/PBE have poor $\omega_0$ and $D_e$, while their
equilibrium bond lengths $r_e$ are within $\sim 2\%$ of experiment.

\section{Summary and Discussion}
\label{sec:summary}

The accuracy of AFQMC depends on the reference wave function $\ket{\Phi_r}$, which is used
to implement the phase-free constraint. \cite{SZ-HK:2003} This is analogous to
DMC, which uses a reference $\ket{\Phi_r}$ to impose the
fixed-node approximation to control the sign problem.
In previous applications, AFQMC was found to have less reliance on the quality of $\ket{\Phi_r}$,
and frequently a single-determinant $\ket{\Phi_r}$ was found adequate.
In these cases, the best results were obtained using
the best variational single determinant reference state, namely
the HF solution when RHF and UHF are the
same (e.g., in the {\HzO} molecule at
equilibrium\cite{Al-Saidi_GAFQMC:2006}), or the UHF
solution $\ket{\Phi^{\textrm{UHF}}_r}$ when the two differ. Moreover, the AFQMC method seemed
relatively insensitive, within the spin unrestricted framework, to whether
a HF, DFT, or hybrid B3LYP Slater determinant was used as $\ket{\Phi_r}$.\cite{Al-Saidi_Post-d:2006}
In some cases, however, such as the BH and N$_2$ molecules, achieving quantitative accuracy
of a few $\mEh$ for the entire PEC
required multi-determinant $\ket{\Phi_r}$.
\cite{Al-Saidi_Bondbreak:2007}

It is shown here that, with $\ket{\Phi^{\textrm{UHF}}_r}$,
the AFQMC PEC of the difficult \Fz{} molecule is
qualitatively incorrect in the intermediate dissociation region.
Spin-contamination is identified as the
primary source of the error.
We have introduced a simple scheme, sp-AFQMC, that
effectively removes spin-contamination effects, regardless of the
choice of $\ket{\Phi_r}$.
It is also illustrated how spin projection can often shorten the
AFQMC equilibration time.
F$_2$ calculations with sp-AFQMC/UHF
were shown to give a PEC whose accuracy is better than a few $\mEh$
across the entire curve in the cc-pVDZ basis.
To our knowledge, these are the most accurate results obtained by
a theoretical method that easily scales up in system size.
The full PEC curves from equilibrium to the dissociation
limit were then calculated with cc-pVTZ and cc-pVQZ basis sets.
Spectroscopic constants with the cc-pVQZ basis were found to be in
excellent agreement with experiment.

The sp-AFQMC results with a single determinant $\ket{\Phi^{\textrm{UHF}}_r}$
are comparable to those obtained with a multi-determinant
$\ket{\Phi_r}$ trial WF from GVB or CASSCF.
The spin-projection method thus further reduces the reliance of AFQMC on the choice of
$\ket{\Phi_r}$, which is one of its most desirable features.

While the focus has been mainly on AFQMC using a single determinant
$\ket{\Phi^{\textrm{UHF}}_r}$ reference wave function, the method may
also prove useful with multi-determinant $\ket{\Phi_r}$ with
significant spin contamination. This could arise, for example, in
treating correlated transition metal systems with truncated CASSCF
wave functions.  
\COMMENTED{
In summary, the spin projection method introduced here
further reduces the reliance of AFQMC on the choice of
$\ket{\Phi_r}$, which is one of its most desirable features.
}

\begin{acknowledgments}

This work was supported by DOE/CMSN (DE-FG02-07ER46366), ONR
(N000140110365 and N000140510055), NSF (DMR-0535529), and ARO (48752PH).
Calculations were performed at the Center for Piezoelectrics by
Design, and the College of William \& Mary's SciClone cluster.
We are grateful to Eric Walter for many useful discussions.
The matrix elements used in our AFQMC calculations were obtained using a
modified {\NWCHEM} 4.6 code. 
The trial wave functions were obtained using {\NWCHEM} and \GAUSSIAN[98]
codes.

\end{acknowledgments}

\bibliography{QMC}

\begin{thebibliography}{42}
\bibitem[{not()}]{note_BrokenSym_HS}
\bibinfo{note}{A spin (as opposed to charge) decomposition
  $(n_{i,\uparrow}-n_{i,\downarrow})$, which breaks spin symmetry, is found to
  be more efficient and is used in most lattice model constrained path Monte
  Carlo calculations (see, e.g., Ref.~\onlinecite{sz-cpmc}). This case will
  require further investigation.}
\end{thebibliography}

\ifthenelse{\equal{\PrePrint}{1}}
{
  \printtables
  \listoffigures

  \pagebreak
  W. Purwanto et al., Fig. 1

  \vskip 1cm
  \noindent
  \includegraphics[scale=.55]{\FIGDIR{}F2-cc-pVDZ-meanfield.eps}\\
  \includegraphics[scale=.55]{\FIGDIR{}F2-cc-pVDZ-corr.eps}

  \pagebreak
  W. Purwanto et al., Fig. 2

  \vskip 1cm
  \includegraphics[scale=.55]{\FIGDIR{}F2-cc-pVDZ-QMC-UHF-spin-contm.eps}

  \pagebreak
  W. Purwanto et al., Fig. 3

  \vskip 1cm
  \includegraphics[scale=.55]{\FIGDIR{}F2-cc-pVDZ-QMC-psiT-try.eps}

  \pagebreak
  W. Purwanto et al., Fig. 4

  \vskip 1cm
  \includegraphics[scale=.55]{\FIGDIR{}F2-cc-pVDZ-spQMC-UHF.eps}

  \pagebreak
  W. Purwanto et al., Fig. 5

  \vskip 1cm
  \includegraphics[scale=.55]{\FIGDIR{}BH-cc-pVDZ-eqlb.eps}

  \pagebreak
  W. Purwanto et al., Fig. 6

  \vskip 1cm
  \noindent
  \includegraphics[scale=.55]{\FIGDIR{}F2-cc-pVTZ-spQMC-UHF.eps}\\
  \includegraphics[scale=.55]{\FIGDIR{}F2-cc-pVQZ-spQMC-UHF.eps}

}
{}

\end{document}